\documentclass[secnumarabic, twocolumn, graphics,floatfix, superscriptaddress,tightenlines, aps, prl]{revtex4-1}
\usepackage[dvips]{graphicx}
\usepackage{dcolumn}
\usepackage{bm}
\usepackage{color}

\begin{document}

\title{Nonlinear optical spectroscopy of indirect excitons in coupled
quantum wells.}
\author{P.~Andreakou}
\affiliation{Laboratoire Charles Coulomb, UMR 5221 CNRS/ Universit\'{e} Montpellier 2,
F-34095, Montpellier, France}
\author{S.~Cronenberger}
\affiliation{Laboratoire Charles Coulomb, UMR 5221 CNRS/ Universit\'{e} Montpellier 2,
F-34095, Montpellier, France}
\author{D.~Scalbert}
\affiliation{Laboratoire Charles Coulomb, UMR 5221 CNRS/ Universit\'{e} Montpellier 2,
F-34095, Montpellier, France}
\author{A.~Nalitov}
\affiliation{Institut Pascal, PHOTON-N2, Universit\'{e} Blaise
Pascal, CNRS, 24 avenue des Landais, 63177 Aubi\`{e}re Cedex,
France.}
\author{N.~A.~Gippius}
\affiliation{Institut Pascal, PHOTON-N2, Universit\'{e} Blaise
Pascal, CNRS, 24 avenue des Landais, 63177 Aubi\`{e}re Cedex,
France.} \affiliation{Skolkovo Institute of Science and Technology,
Skolkovo, Moscow Region, 143025, Russia.}
\author{A.~V.~Kavokin}
\affiliation{Russian Quantum Center, 100, Novaya, Skolkovo, Moscow
Region, 143025, Russia}
\affiliation{School of Physics and
Astronomy, University of Southampton, Southampton, SO17 1BJ, United
Kingdom}
\affiliation{Spin Optics Laboratory, St-Petersburg State
University, 1, Ulianovskaya, St-Peterbsurg, 198504, Russia}
\author{M.~Nawrocki}
\affiliation{Institute of Experimental Physics, University of
Warsaw, $Ho\dot{z}a$ 69, 00-681 Warsaw, Poland}
\author{J.~R.~Leonard}
\affiliation{Department of Physics, University of California at San Diego, La Jolla, CA
92093-0319, USA}
\author{L.~V.~Butov}
\affiliation{Department of Physics, University of California at San Diego, La Jolla, CA
92093-0319, USA}
\author{K.~L.~Campman}
\affiliation{Materials Department, University of California at Santa Barbara, Santa
Barbara, California 93106-5050, USA}
\author{A.~C.~Gossard}
\affiliation{Materials Department, University of California at Santa Barbara, Santa
Barbara, California 93106-5050, USA}
\author{M.~Vladimirova}
\affiliation{Laboratoire Charles Coulomb, UMR 5221 CNRS/ Universit\'{e} Montpellier 2,
F-34095, Montpellier, France}

\begin{abstract}
Indirect excitons in coupled quantum wells are long-living
quasi-particles, explored in the studies of collective quantum
states. We demonstrate, that despite the extremely low oscillator
strength, their spin and population dynamics can by addressed by
time-resolved pump-probe spectroscopy. Our experiments make it
possible to unravel and compare spin dynamics of  direct excitons,
indirect excitons and residual free electrons in coupled quantum
wells. Measured spin relaxation time of indirect excitons exceeds
not only one of direct excitons, but also one of free electrons by
two orders of magnitude.
\end{abstract}

\maketitle

%
An exciton is a semiconductor quasi-particle formed by an electron
and a hole bound by Coulomb interaction. A spatially indirect
exciton (IX) can be formed in coupled quantum wells (CQWs) when an
electron and a hole are confined in different quantum wells.
Remarkable features of IX gases, including spontaneous coherence and
condensation
\cite{HighNature,HighPRL2013,DubinEvidencearXiv:1304.4101v1},
long-range spin currents and spin textures
\cite{HighPRL2013,DubinEvidencearXiv:1304.4101v1}, pattern formation
\cite{DubinEvidencearXiv:1304.4101v1,Butov2002,Butov2004,SternScience},
 and correlation phenomena \cite{Remeika2009,Rapaport} have been recently
 demonstrated.
The spin structure of indirect excitons is particularly interesting
and important.
The exciton states with spin projections $\pm 1$ to the structure
axis can couple to light to form bright states, while excitons with
$\pm 2$ spin projections are dark, and nontrivial to access by
emission spectroscopy \cite{MaillaleSham}. Measuring the density and
polarization state of both dark and bright components of the IX gas
is one of challenges in the physics of IX. Another issue is the effect of the residual
two-dimensional electron gas (2DEG) in the gated CQWs on the exciton
spin dynamics \cite{Butov2004,Rapaport2004}.
Indeed, the 2DEG can significantly affect IX spin dynamics and other properties \cite%
{Berman}. Traditional photoluminescence experiments fail in
accessing a dilute 2DEG, while the methods of nonlinear
time-resolved spectroscopy, such as pump-probe Kerr rotation and
reflectivity, may bring new insights. Pump-probe techniques have
been successfully applied to investigate electron
\cite{Kikkawa1997}, hole \cite{Syperek}, direct exciton (DX)
\cite{Portella-Oberli2004,Malinowski} and exciton-polariton
\cite{Brunetti2006} dynamics in nanostructures. But their
application to IX spectroscopy is limited, because of the extremely
low IX oscillator strength.
 Due to this fundamental reason, studies of indirect
excitons have been limited to linear optics methods until now. 
%

\begin{figure}[t]
\center{\includegraphics[width=1\linewidth]{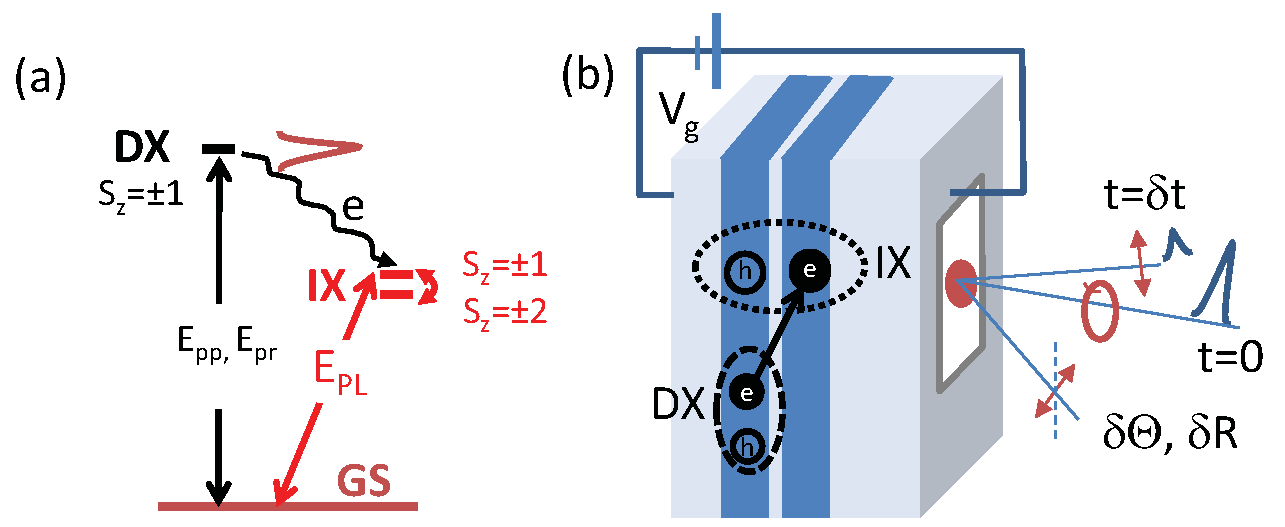} } \caption{
(a) Three-levels scheme of the pump-probe experiment with relevant
excitonic states in a biased CQWs. Low oscillator strength IX states
are pumped ($E_{pp}$) and probed ($E_{pr}$) via DX transition,
through their common ground state. (b) Sketch of CQWs and pump-probe
experiment. } \label{setup}
\end{figure}

\begin{figure}[t]
\center{\includegraphics[width=1\linewidth]{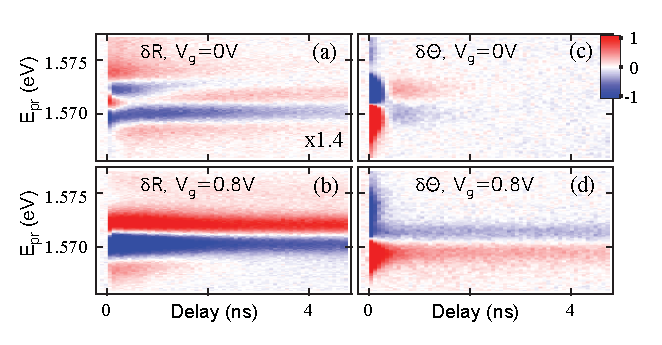} } \caption{
Reflectivity (a, b) and Kerr rotation (c, d) induced by pump pulses
resonant with DX transition ($E_{pp}=1.568$ eV), measured as a
function of the pump-probe delay and probe energy at $V_g=0$ (a, c)
and $V_g=0.8$~V (b, d). } \label{decay}
\end{figure}

In this Letter we report a proof-of-concept experiment demonstrating
the potential of time-resolved pump-probe spectroscopy of IXs in
CQWs. The  method is based on the three-level scheme (Fig.
\ref{setup}~(a)). In this scheme pump and probe light pulses are
resonant with the optically active DX transition, and both bright
and
 dark IXs are probed via their ground state, common with the DXs. The
experiment reported here is the first realization of our theoretical
proposal \cite{Nalitov2013}.
%
We show that both DX and IX spin and population dynamics, as well as the
spin polarization of residual electrons may be detected via the modulation
of reflectivity and Kerr rotation spectra at the DX resonances. From the
photoinduced reflectivity measured as a function of the pump-probe delay we
extract DX and IX radiative lifetimes varying from $1$ to $30$ ns, with a
clear footprint of the gate voltage controlled DX-IX anticrossing. %
%
In the Kerr rotation signal, we unravel DX, IX and electron spin
dynamics. In unbiased CQWs electron spin contribution is negligible,
and the coherent dynamics of DXs dominates. In biased CQWs the
electron density and spin polarization build up. It appears that not
only DX, but also bare electron spin relaxation is much faster
($\simeq 200$~ps) than the spin relaxation of IXs (up to $10$ ns).
We show that this is due to more efficient localization of IXs. From
the spectral shape analysis of the nonlinear signals we conclude
that the main mechanisms of IX-DX interaction are the
spin-independent narrowing of the DX resonance and the strongly spin
dependent blue shift of the DX energy.

The sample we study consists of two $8$ nm wide GaAs quantum wells separated
by a $4$ nm Al$_{0.33}$Ga$_{0.67}$As barrier and surrounded by $200$ nm Al$%
_{0.33}$Ga$_{0.67}$As layers. The top and bottom electrodes are
n-GaAs layers. The voltage $V_{g}$ applied between the conducting
n-GaAs layers drops in the insulating layer between them. The
details on this sample can be found in Ref. \cite{Butov1999}. In all
experiments the sample is immersed in superfluid helium. We perform
Kerr rotation and photoinduced reflectivity experiments, in which
spin-polarized DXs are optically excited in the CQW by a circularly
polarized pump pulse, Fig.~\ref{setup}~(b). The resulting dynamics
of the spin polarization (density) is monitored via Kerr rotation
(reflectivity) of the delayed linearly polarized probe pulse.
Two-color measurements are realized by spectral filtering of pump
and probe pulses with two $4f$ zero-dispersion lines. The pulse
duration is $1$ ps, the spectral width is $1.5$ meV. The Ti-Sapphire
laser repetition rate is reduced to $20$~MHz by a pulse-picker in
order to avoid exciton accumulation between pulses. The laser spot
diameter on the sample is $100$~$\mu $m, typical powers are $120$
and $70$ $\mu $W for pump and probe, respectively \footnote{We have
checked that reducing probe to pump power ratio does not change the
signal dynamics.}.

The photoinduced reflectivity signal $\delta R$ provides information
on the total exciton density in the CQWs.
Fig. \ref{decay}~(a, b) shows  $\delta R$ measured as a function of
the pump-probe delay $\delta t$ and probe energy $E_{pr}$, at fixed
pump energy $E_{pp}=1.568$~eV.
These data show that even in the unbiased device the photoinduced
reflectivity, and thus the exciton population persist as long as
$5$~ns. At short pump-probe delays, a double resonance structure is
apparent.
%
In the biased structure (Fig. \ref{decay} (b)) the spectral profile
of $\delta R$ also changes significantly during the first nanosecond
after the pump pulse. However, a strong signal persists at the
longest delays studied. Decreasing the time interval between laser
pulses from $48$~ns to $24$~ns  leads to the accumulation of the
excitons between pulses and  the pump-probe signal at negative
delays builds up. This means, that the life-time of excitons in this
structure is of the order of $30$~ns, consistent with the IX
photoluminescence kinetics measurements \cite{Butov1999}.

\begin{figure}[tbp]
\center{\includegraphics[width=1\linewidth]{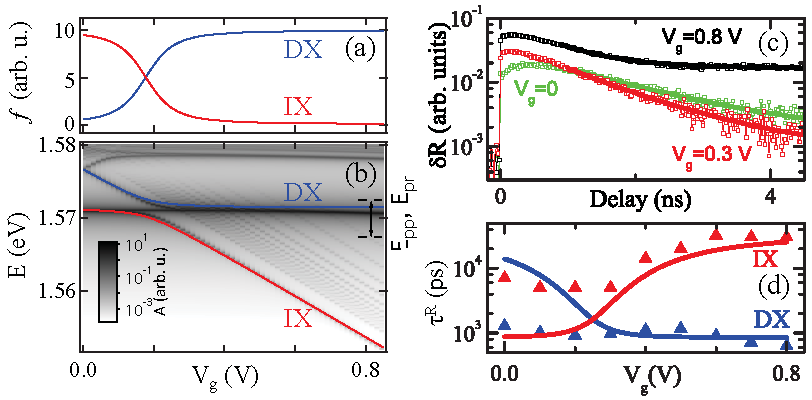} } \caption{(a)
Oscillator strengths and (b) energies of DX and IX calculated from
the coupled oscillators model as a function of the gate voltage.
Color map in (b) shows accurate solution for excitonic absorbtion
obtained from Schr\"{o}dinger equation
\protect\cite{Sivalertporn}.(c) Photoinduced reflectivity measured
(symbols) as a function of $\delta t$ at $E_{pr}=1.570$~eV, solid
lines are fit to bi-exponential decay. Pump energy $E_{pp}=1.568$
eV. (d) IX and DX lifetimes extracted from bi-exponential fit of the
reflectivity decay (symbols), and inverse oscillator strengths of DX
and IX coupled oscillators (solid lines).} \label{figR}
\end{figure}

Fig. \ref{figR}~(c) shows typical time scans of the photoinduced
reflectivity at three different values of the gate voltage and fixed
probe energy, which show non-monotonous behavior. This
non-monotonous behavior  is due to the shift of the spectrum during
the first nanosecond after the pump pulse (Fig. \ref{decay}~(a, b)).
At longer delays the decay becomes bi-exponential, with the
characteristic times plotted in Fig. \ref{figR}~(d) as a function of
the gate voltage. While the shortest decay time $\sim 1$~ns does not
depend on the voltage, the longer decay time has a pronounced
voltage dependence. The decay of the excitonic population is faster
at $V_{g}=0.3$~V than the decay in the unbiased device.

The  observed dynamics of the exciton population, as well as its
voltage dependence  can be understood as a footprint of both DXs and
IXs, and their voltage-controlled anticrossing.
Indeed, the exciton states in CQWs are formed from four possible
electron-hole pair states. The corresponding exciton energies and
oscillator strengths  can be accurately calculated by solving
Schr\"{o}dinger equations \cite{Sivalertporn}. The resulting DX-IX
anticrossing is well described by the simple two-level anticrossing
picture (Fig. \ref{figR} (a, b)).
At  $V_{g}=0$, DX is the ground state of the system, IX is only
several meV above it, and its oscillator strength is only $10$ times
smaller.
In this case $\delta R$ spectral shape exhibits the double resonance
structure, with both DX and IX contributions (Fig. \ref{decay}~(a)),
and two decay times which differ by a factor of $10$ (Fig.
\ref{figR} (d)).
By contrast, at $V_{g}=0.8$~V, the IX state is about $16$~meV below
the DX and has an oscillator strength $100$ times smaller than DX.
Thus, the fast decay of the $\delta R$  signal is due to DX
recombination, and  slow relaxation is due to IXs recombination
(Fig. \ref{figR} (d)).
Remarkably, neither pump nor probe pulses are resonant with IX
transition, so that the entire nonlinear signal results from
DX-IX interaction.
%
%
The smallest difference between the two decay times is observed at
$V_{g}=0.3$~V, close to the DX-IX anticrossing point.
The gate voltage dependence of the two decay times fits to the
coupled oscillator model shown in Fig. \ref{figR}~d.
The complex spectral shape and dynamics of $\delta R$ at short
pump-probe delays is due to the high density of both DXs and IXs in
the device, its detailed understanding is beyond the scope of this
work.

\begin{figure}[tbp]
\center{\includegraphics[width=1\linewidth]{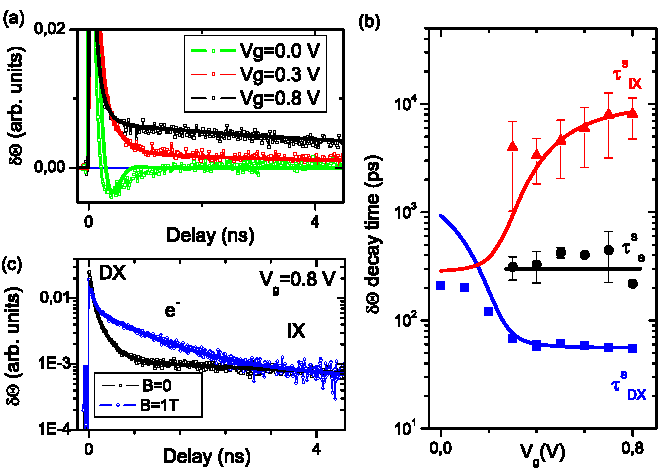} } \caption{
(a) Photoinduced Kerr rotation (symbols) as a function of
$\delta t$ at $E_{pr}=1.569$~eV, solid lines are damped cosine (%
$V_g=0$) and triple exponential decay ($V_g=0.3$ and $0.8$~V) fitted
curves.
Pump energy E$%
_{pp}=1.568$ eV. (b) Spin lifetimes extracted from Kerr rotation
decay as a function of gate voltage (symbols). Solid lines: coupled
oscillators model for DX and IX, estimation from Dyakonov-Perel
mechanism for electrons. (c) Same as (a) in log scale at
$V_g=0.8$~V, $B=0$ (black squares) and at $B=1$~T (blue circles).
Solid lines are triple exponential decay fit to the data.}
\label{figK}
\end{figure}

Kerr rotation measurements provide the information on the spin
dynamics in CQWs and the results are even more intriguing (Fig.
\ref{decay}~(c, d)). At $V_{g}=0$ the rapid decay of $\delta \Theta$
is accompanied by the inversion of the derivative-like spectrum at
$0.5$~ns. In contrast, at $V_{g}=0.8$~V the rapid decay is followed
by much slower relaxation on the scale of several nanoseconds
without the inversion of the spectrum. In Fig. \ref{figK}~(a) we
plot the Kerr rotation  measured at fixed probe energy. At $V_g=0$
the signal is nonmonotonous, consistent with the spectrally resolved
measurements of Fig.\ref{decay}~(c). At higher voltages, the decay
is triple exponential, with the longest spin lifetime reaching
$10$~ns. The decay times obtained by fitting the data at different
gate voltages are summarized in Fig. \ref{figK}~(b).

We propose the following interpretation of these observations. First
of all, in semiconductor heterostructures a small splitting between
two perpendicularly polarized linear exciton states $\delta _{xy}$
is generally present \cite{HighPRL2013}. The optical birefringence
due to this splitting has been studied in detail in QW microcavity
structures \cite{Brunetti2007}. For an exciton spin this splitting
acts as an effective in-plane  magnetic field. Therefore, relaxation
of the spin polarization is accompanied by its rotation around this
effective field with an angular frequency $\Omega _{xy}=\delta
_{xy}/\hbar $, provided that the precession is fast enough with
respect to the exciton spin relaxation $\Omega _{xy}>1/\tau _{s}$.
Indeed, Kerr rotation at zero and small bias is well described by
the damped cosine function (solid line), assuming the Gaussian
distribution of precession frequencies centered at $\omega =6$~GHz
(corresponding to $\delta _{xy}=4$~$\mu $eV), with the standard
deviation $\sigma =\pm 2.5$~GHz, and the exciton spin relaxation
time $\sim 200 $ ~ps. The latter is longer than DX spin relaxation
expected in a single QW of the same width
\cite{DamenAPL1991,MaillaleSham}. The DX coupling to the
energetically close IX state is probably responsible for this
effect.
Coherent rotation of the exciton, rather than an electron spin in
the presence of the applied in-plane magnetic field have been
observed in time-resolved PL experiments \cite{Amand1997}. It was
shown to be governed by the stability of the hole spin within the
exciton. The hole spin relaxation time $\tau _{s}^{h}$ should
satisfy the condition $1/\tau _{s}^{h}<\delta _{exc}/\hbar$, where
$\delta _{exc}$ is the electron-hole exchange energy
\cite{Dyakonov1997}.
 Assuming $\delta _{exc}=70$~$\mu $eV in the unbiased
structure \cite{Blackwood1994} yields $\tau _{s}^{h}>10$ ps, which should be
fulfilled in the studied structure under a low power excitation \cite%
{Fokina2010}. Application of the electrical bias reduces the
exchange energy \cite{Vina}. Indeed, at $V_{g}>0.2$~V the Kerr
rotation decay becomes monotonous.
At high voltages, the shortest decay  time of order of $50$~ps can
be attributed to the polarization relaxation of DXs
\cite{DamenAPL1991,MaillaleSham}. The two other components have
characteristic decay times of the order of $200$~ps and several
nanoseconds, respectively, the latter increasing with the gate
voltage. They can be attributed to the spin relaxation of the 2DEG,
which forms in biased CQWs \cite{Butov2004,Rapaport2004,ButovJETP},
and IXs.

One of the main results of this work is the distinction between the
spin polarized IXs and the spin polarized 2DEG traditionally studied
by pump-probe Kerr rotation technique.
The voltage dependence of the spin dynamics suggests, that fastest
and slowest components are due to DX and IX, respectively. Indeed,
we reproduce the voltage dependence of these two components within
the model of the two coupled oscillators (solid lines in Fig.
\ref{figK}~(b)). By contrast, the electron spin relaxation is not
expected to be voltage dependent, the Rashba contribution to the
spin-orbit field being small with respect to the
Dresselhaus field \cite{KavokinPortnoi,Studer2009}.  Within IXs we
deal with the spin relaxation of an electron bound to the hole, but
their exchange coupling is negligible at high $V_g$.
The key
parameter which controls the spin relaxation of electrons is the
degree of their localization in the disorder potential
\cite{Cundiff}.
Indeed, the electron spin relaxation time due to spin-orbit coupling
can be written as  $\tau_{s}^{e}=(1+\tau _{0}/\tau _{c})/(\tau _{c}
\Omega _{SO}^{2})$ \cite{SM}.
Here $\tau_0$ is the characteristic time during which an electron remains localized
and is not affected by the spin-orbit field,
$\tau_c$ is the correlation time of the spin-orbit field \footnote{Correlation time of a fluctuating field is a
 time during which the field may be considered as constant. In the case of the
 spin-orbit field and mobile electrons it's given by the momentum scattering time,
 while in the hoping regime  it's rather the time of flight.}, and $\Omega _{SO}$ is the spin-orbit
 frequency.
Depending on the  relative values of $\tau _{c}$ and $\tau _{0}$,
elections (IXs) are either localized  $(\tau _{c}  \ll \tau _{0})$,
or mobile $(\tau _{c}  \gg \tau _{0})$.
One can easily see that faster spin relaxation is expected for
mobile, than for localized electrons \cite{SM,Dzhioev2002,Cundiff}.
 %
%
Because an exciton is  heavier than an electron, at low density it
can be efficiently localized by the disorder potential, while at
high density the mobility threshold can be reached
\cite{Leonard2009}.
To check this idea, we studied the power dependence of the Kerr
rotation. Increasing the pump power allows to increase the IX
density up to the mobility threshold, while keeping the electron
density fixed. It turns out that the slow component disappears above
a critical power \cite{SM}. This corroborates the interpretation of the slow
component in terms of the spin polarization of the localized IX,
which disappears above the mobility threshold.
 On the other hand, increasing excitation power results to a
 relatively weak gradual variation of the fast component that supports its
 assignment  to the residual electron spin.
Electron spin relaxation is accelerated due to the heating of the
electron gas by the photogenerated carriers \cite{Zhukov2007}.

The ultimate  test of this interpretation is the application of a
longitudinal magnetic field, which suppresses spin relaxation of
residual electrons much more efficiently, than that of the electrons
bound to holes within IXs \cite{SM}.
For electrons within IX the suppression factor is given by
$\tau_{s}^{e,B}/\tau_{s}^{e}=1+(\Omega _{L}\tau _{c})^{2}$, where
$\Omega _{L}$ is the Larmor frequency \cite{DyakonovBook}.
For the electrons which are not bound into the excitons and thus are
subject to the cyclotron motion, $\tau _{s}^{e,B}/\tau
_{s}^{e}=1+(\Omega _{c}\tau _{c})^{2}$ , where $\Omega _{c}$ is the
cyclotron frequency.
Because the cyclotron frequency is usually higher than the Larmor
frequency, spin relaxation of the free electrons is expected to be
much strongly
 affected by the magnetic field.
 Moreover, since the same $\tau _{c}$
 is involved in the spin relaxation and it's quenching in the presence of the
 magnetic field,  measuring both times further tightens the conditions
 on the localization degree
 of both excitons and electrons.
 Fig.~\ref{figK}~(c) shows Kerr rotation at $V_{g}=0.8$~V as a function of
the pump-probe delay at $B=0$ and $B=1$~T. Among the three decay
components, only one with $\tau _{s}^{e}=200$~ps is affected by the
magnetic field. A systematic study of the  magnetic field effect
 shows that it increases by a
factor of three for all $V_{g}>0.3$~V, as expected for the electrons,
while slow component remains unchanged, as expected for IXs.
The  analysis of these results in the framework of the spin orbit
relaxation model is reported in the Supplemental Material.
It allows for the identification of the  fast decay as being due to
the spin relaxation
 of resident electrons, and
 the  slow decay as being due to the
localized IXs.
$\tau _{s}^{IX}$ reaches about $10$~ns (Fig.
\ref{figK} (c)), consistent with the polarization lifetime of
localized IXs measured in PL
imaging experiments \cite{Leonard2009}. %

\begin{figure}[tbp]
\center{\includegraphics[width=1\linewidth]{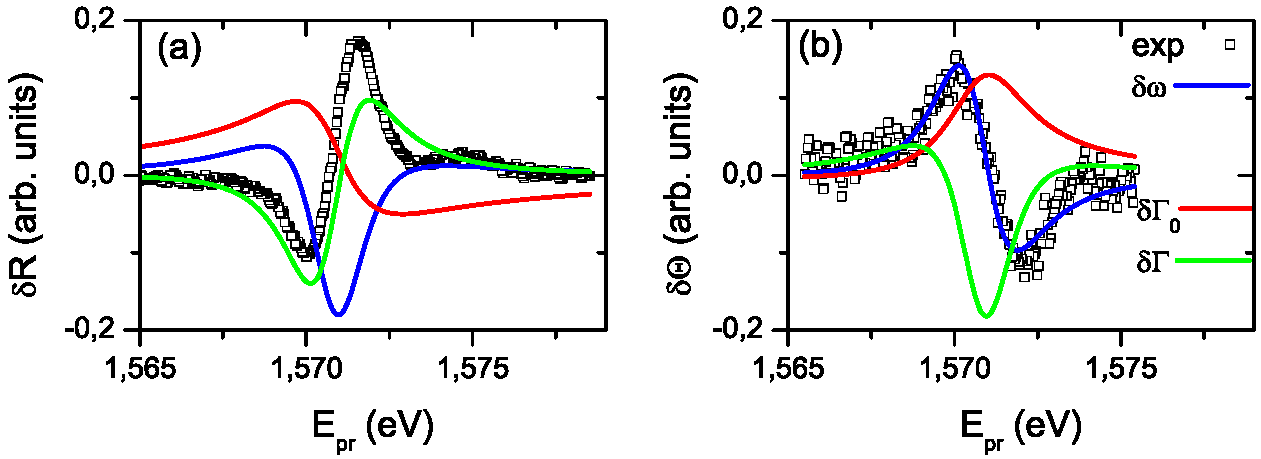} }
\caption{Probe spectra of photoinduced reflectivity (a) and Kerr
rotation (b) measured at $5$~ns pump-probe delay, $E_{pp}=1.568$~nm
(squares). Lines are calculated assuming photoinduced modification
of the DX properties: energy blue shift (blue), saturation (red) and
narrowing (green) \cite{Nalitov2013}.} \label{spectra}
\end{figure}

%
Spectral analysis of the $\delta R$ and $\delta \Theta$
provide important information on the excitonic nonlinearities. %
%
We focus here on the behavior at high voltages and long pump-probe
delays, where the signal is dominated by IXs. Fig. \ref{spectra}
shows the spectra measured at $V_{g}=0.8$~V and $\delta t=5$~ns. To
identify the dominant nonlinearity we follow the roadmap proposed in
Ref. \cite{Nalitov2013}. First of all we measure linear reflectivity
in the vicinity of the DX resonance.  The presence of IXs can modify
the linear reflectivity via one of the three mechanisms: energy
shift $\delta \omega$, saturation $\delta \Gamma_0$ and
narrowing/broadening of the resonance $\delta \Gamma$
\cite{Fokina2010,Nalitov2013,CundiffPhysicaE}. The spin independent
mechanisms contribute to the photoinduced reflectivity only, while
spin-dependent ones are responsible for the Kerr rotation. Fitting
to the data in Fig.~\ref{spectra}~(a) shows that the photoinduced
reflectivity signal is dominated by the DX resonance narrowing. This
can be understood as a reduction of the DX inhomogeneous broadening
due to the screening of the QW disorder potential by localized IXs.
Thus, at least at low IX densities, the impact of the DX blue shift
due to the IX-DX interaction on the pump-probe spectra is weaker
than that of the narrowing of the DX line.
However, the blue shift of the DX resonance provides the main
contribution to the Kerr rotation, Fig.~\ref{spectra}~(b).
This is consistent with a strong spin dependence of the DX blue shift \cite{Nalitov2013}.
%

%

In conclusion, we have shown that despite their vanishing oscillator
strength, IXs in biased CQWs can be efficiently addressed by
pump-probe spectroscopy. The detection of both photoinduced
reflectivity and Kerr rotation provides a powerful tool for
unraveling the spin dynamics of IXs and the 2DEG, exploring IX-DX
interaction and probing both bright and dark IX populations. In the
appropriately chosen CQW devices, this method may help solving the
challenging problems of the exciton physics, which are not easily
accessible by other experimental means, such as determination of
relative density and spin polarization of the bright and dark IX
states.

\emph{Acknowledgments.} We are grateful to K.~V.~Kavokin and M. I. Dyakonov for
valuable discussions and acknowledge the support of EU ITN INDEX
PITN-GA-2011-289968. LVB  was supported by DOE Award
DE-FG02-07ER46449 and  JRL by a Chateaubriand Fellowship. AK
acknowledges the support from the Russian Ministry of Education and
Science, Contract No 11.G34.31.0067. MN acknowledges the support
from the Polish National Science Center under decision
DEC-2013/09/B/ST3/02603.


\pagebreak \widetext
\begin{center}
\textbf{\large Supplemental Material}
\end{center}
\setcounter{equation}{0} \setcounter{figure}{0}
\setcounter{table}{0} \setcounter{page}{1} \makeatletter
\renewcommand{\theequation}{S\arabic{equation}}
\renewcommand{\thefigure}{S\arabic{figure}}

\section{Dependence of the indirect exciton spin and population on the pumping energy and power.}
\begin{figure}[tbp]
\centering
\includegraphics[width=0.5\linewidth]{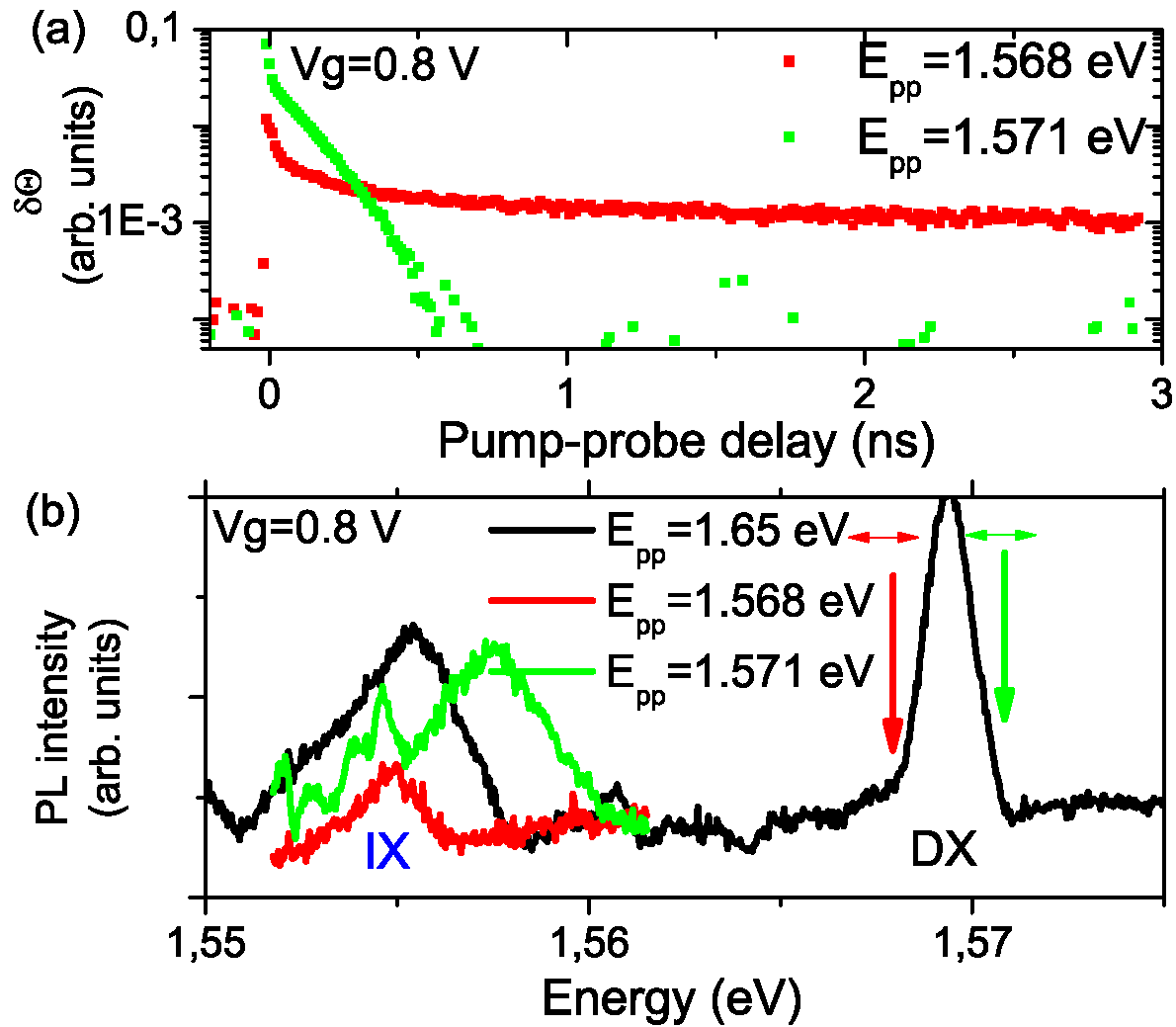}
\caption{(a)  Photoinduced Kerr rotation  as a function of
pump-probe delay measured at $E_{pr}=1.569$~eV, $V_g=0.8$~V. Pump
energies are E$_{pp}=1.568$~eV (red) and E$_{pp}=1.571$~eV (green).
(b) Photoluminescence spectra measured at $V_g=0.8$~V for different
excitation energies $E_{pp}$. IX  emission is blue shifted at
E$_{pp}=1.571$~eV (green) with respect to E$_{pp}=1.568$~eV (red)
and non-resonant excitation E$_{pp}=1.65$~eV (black).}
\label{GraphPLKerr}
\end{figure}

%
%
In this section we show that at given pumping power, the spin
polarization of indirect excitons (IX) is optimized for the
excitation energy slightly below DX resonance, and that this can be
understood in terms of the density dependence of the exciton spin
relaxation.
Fig. \ref{GraphPLKerr} (a) shows Kerr rotation measured at
$V_g=0.8$~V for two different excitation energies and same power
$P=120$~$\mu$W as the measurements shown in the main text.
Probe energy is chosen to optimize the signal at $3$~ns pump-probe
delay.
One can see, that under high energy pumping slow component of the
Kerr signal disappears completely, so that no signal related to the
IX spin dynamics can be identified.
The energy of the pump beam with respect to the non-resonantly
excited photoluminescence (PL) of  the CQW device at the same gate
voltage (black line) is shown in Fig. \ref{GraphPLKerr}~(b).
One can see, that the two pump energies are situated on the
different sides of the DX emission line.
For roughly $1$~meV shift between DX absorbtion and  emission and
$1.571$~eV corresponding to the absorption maximum \cite{ButovJETP},
we argue that at $E_{pp}=1.571$~eV resonant excitation maximizes the
IX density, while at $E_{pp}=1.568$~eV the IX density is much lower.
This is unambiguously confirmed by resonant PL experiments for this
two pump energies (Fig. \ref{GraphPLKerr}~(b), green and red lines).
The PL measurements are taken in exactly the same conditions as the
pump-probe measurements.
For the high energy excitation the IX emission is $3$~meV above the
IX emission at low energy excitation.
The blue shift of the emission is a clear signature of the higher IX
density $n_{IX}$ and the resulting delocalization of IX, because the
amplitude of the disorder potential is expected to be of order of
$1$~meV.
In the framework of the plain capacitor model, the IX-IX interaction
energy is given by $u_0 n_{IX}$, where  $u_0=4\pi e^2 d/ \epsilon$,
$d$ is the separation between the QWs, $e$ is the electron charge,
$\epsilon$ is the background dielectric constant
\cite{IvanovEL2002}.
From the measured blue shift  this approximation gives for the
studied structure under high energy excitation $n_{IX}\simeq 2 \cdot
10^{10}$~cm$^{-2}$, and not more than $5 \cdot 10^{9}$~cm$^{-2}$ at
low energy excitation.
%
%
\begin{figure}
\centering
\includegraphics[width=0.5\linewidth]{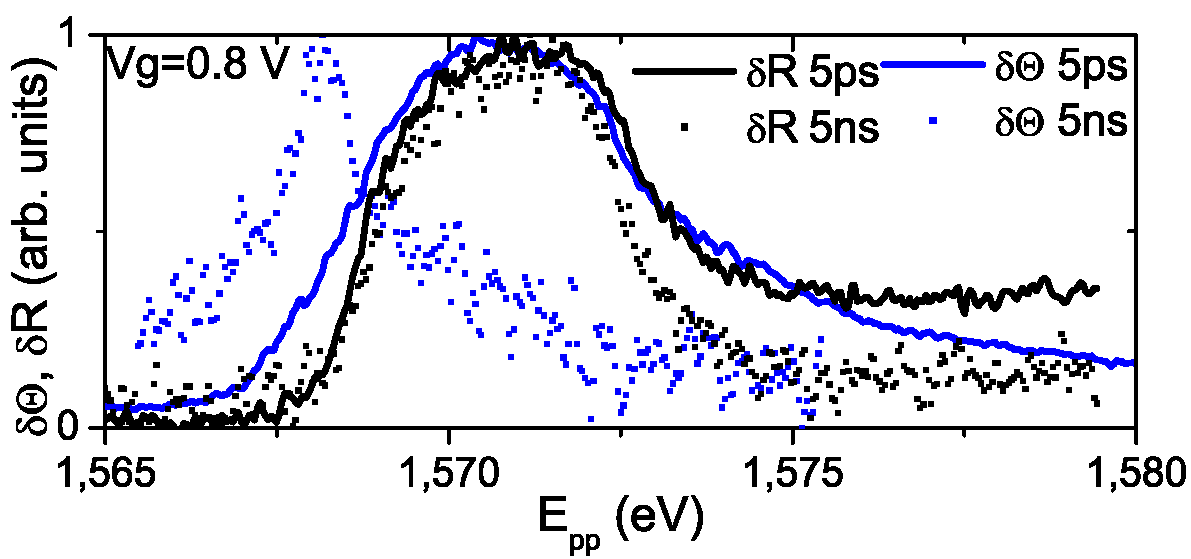}
\caption{Normalized Kerr rotation (blue) and photoinduced
reflectivity (black) excitation spectra measured at E$_{pr}=1.57$~eV
and $V_g=0.8$~V. Short pump-probe delay ($5$~ps, line) and long
pump-probe delay ($5$~ns, symbols) are compared.
}\label{GraphPumpSpectra}
\end{figure}

%
In order to confirm further the role of the exciton density in the
spin relaxation processes, we report in Fig. \ref{GraphPumpSpectra}
the normalized excitation spectra of the  Kerr rotation and
reflectivity measured immediately after the excitation (pump-probe
delay $\delta t=5$~ps) and at long delays $\delta t=5$~ns.
The probe energy was fixed at $E_{pr}=1.57$~eV.
One can see from the measurements at short delays, that the same
pumping energy optimizes the carrier density ($\propto \delta R$)
and the spin density ($\propto \delta \Theta$) created by the pump
pulse.
By contrast, at $\delta t=5$~ns the spin polarization is a trade off
between the initial polarization and the spin relaxation rate.
The spin polarization maximum is achieved at lower energy than the
maximum of the IX population.

\begin{figure}
\centering
\includegraphics[width=0.5\linewidth]{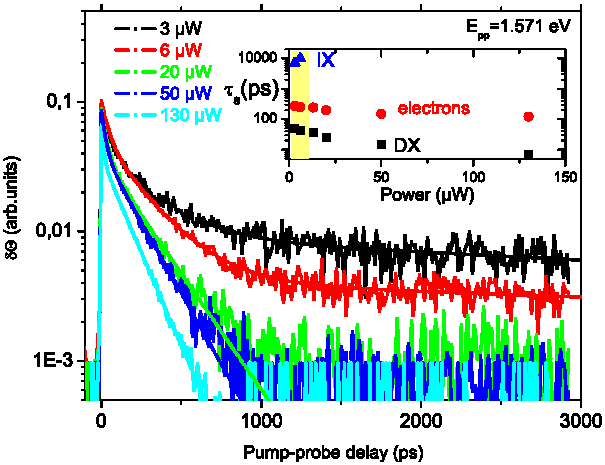}
\caption{Kerr rotation (symbols) measured as a function of the
pump-probe delay at $E_{pr}=1.57$~eV, $V_g=0.8$~V. At typical pump
power $P_{pp}=130$~$\mu$W used throughout this work no long-living
spin polarization is detected at the pump energy $E_{pp}=1.571$~eV.
Lowing down the pump power allows to  recover the slow component.
Solid lines are fits to bi-exponential ( $P_{pp}=130$ to
$P_{pp}=20$~$\mu$W) or tree-exponential ( $P_{pp}=6$ and
$P_{pp}=3$~$\mu$W) decay.  Inset: spin life-times of DXs, electrons
and IXs extracted from the fits. The spin life-times in the yellow
region are similar to the results obtained at $E_{pp}=1.568$~eV and
$P_{pp}=130$~$\mu$W, described in the main text.}
\label{GraphPowerDep}
\end{figure}
%
Finally, it's important to check, that lowing down the excitation
power at $E_{pp}=1.571$ eV allows to recover the localization and
slow spin relaxation of IXs.
Fig. \ref{GraphPowerDep} shows power dependence of the Kerr rotation
signal at $V_g=0.8$~V, $E_{pp}=1.571$~eV, solid lines are fits to
bi- or three-exponential decay.
With decreasing power, both electron and DX spin relaxation slow
down progressively, but the slow component related to IX builds up
only below the critical power (see inset of Fig.
\ref{GraphPowerDep}).
This power should correspond to the IX density at which the IX-DX
interaction energy is smaller than the disorder potential amplitude
\cite{Remeika2009}.
Thus we conclude that  IXs spin dynamics is mainly determined by its
localization degree, Dyakonov-Perel spin relaxation being abruptly
quenched for the excitation powers below IX localization transition.
%

\begin{figure}
\includegraphics[width=0.5\linewidth]{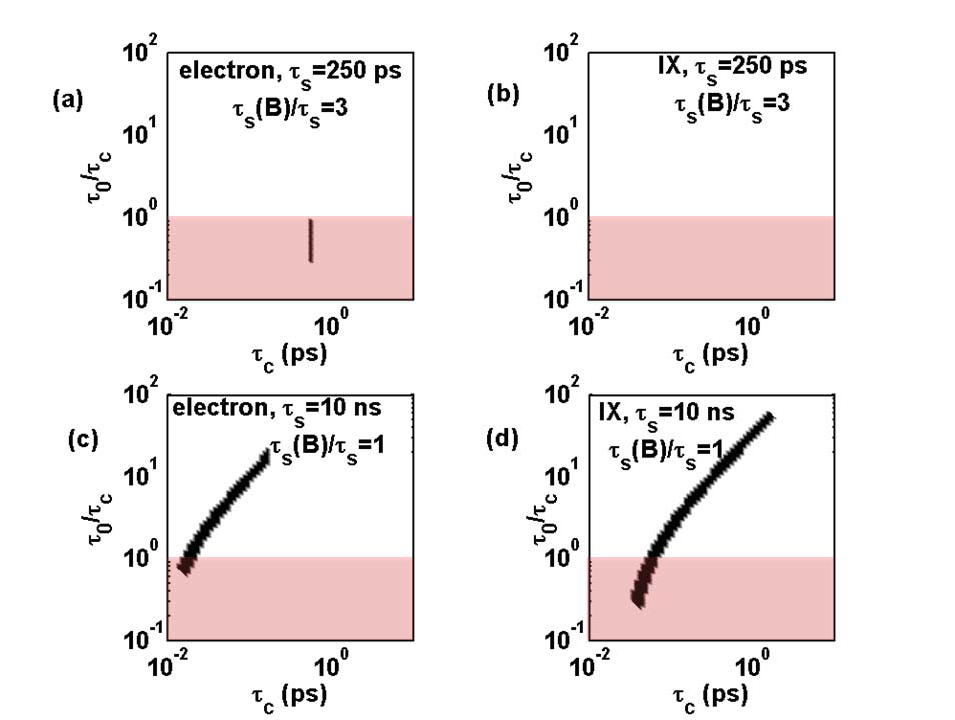}
\caption{The two components of spin polarization signal mapped on
($\tau_c$, $\tau_0/\tau_c$) parameter space. Pink areas denote the
mobility area where $\tau_c > \tau_0$. Fast component
($\tau_s=250$~ps), $\tau_s(B)/\tau_s=3$) is calculated assuming
either electron (a) or IX (b) spin relaxation. There are no points
in the parameter space for the fast relaxation of IX. Slow component
($\tau_s=10$~ns), $\tau_s(B)/\tau_s=1$) can be obtained for both
electrons (c) and excitons (d).}
 \label{FigTau}
\end{figure}

\section{Model for electron and IX spin relaxation.}
The key parameter which controls the spin relaxation of electrons is
the degree of their localization in the disorder potential
\cite{Cundiff}.
First of all in the regime of the strongest localization the spin
relaxation of electrons is due to hyperfine interaction with a
limited number of nuclear spins, which acts on the electron as a
fluctuating effective magnetic field.
This type of relaxation is typical for electrons localized in
natural quantum dots, the characteristic spin decay time are on the
scale of nanoseconds, and this relaxation is efficiently suppressed
by the longitudinal magnetic field of the order of mT
\cite{Merkulov02,Gerlovin2007}.
In the experiments presented in this work long-living component of
the spin polarization does not present any field dependence.
Hence, we conclude that hyperfine field is not the main source of
the spin relaxation in our CQWs, and will neglect the spin
relaxation of localized electrons (IXs).

The other source of relaxation is the fluctuating effective magnetic
field due to spin-orbit interaction.
While an electron (or IX) hops from one localization site to
another, its spin experiences random rotations.
In bulk zinc blende crystals this mechanism is a major cause of spin
relaxation for the localized states in the impurity-band
\cite{KavokinSST09}.
In order to take into account the localization, we  describe the
diffusion of an electron (IX) in the QW in-plane disorder potential
by two times.
These are correlation time of the fluctuating spin-orbit field
$\tau_c$, during which the spin-orbit field can be considered as
constant, and $\tau_0$, during which an electron remains localized
and not affected by the spin-orbit field.
The strength of the spin-orbit interaction is given by the root mean
square value of the electron Larmor frequency in the spin-orbit
field $ \Omega _{SO}=2\beta k/ \hbar$, where
 $\beta =2$~$\mu$eV$\cdot\mu $m is the spin-orbit constant
\cite{KavokinPortnoi,Leonard2009} and $k$ is the wave vector of the
electron (IX).
We shall assume that both residual electrons and electrons bound to
holes within photoinduced IXs are characterized by their thermal
wave vectors, $k_e=17$~$\mu$m$^{-1}$, $k_{IX}=10$~$\mu$m$^{-1}$ at
$T=2$~K.
If the in-plane motion of an electron (IX) consist in hoping between
the localization sites, the well-known Dyakonov-Perel formula for
the spin relaxation time \cite{DyakonovBook} should be scaled by the
factor $1+\tau_0/\tau_c$, so that  $\tau_{s}^{e(IX)}=(1+\tau
_{0}/\tau _{c})/(\tau _{c} \Omega _{SO}^{2})$.
%
In the limit of free carriers ($\tau _{c} \gg \tau _{0}$)  this
formula is reduced to the usual Dyakonov-Perel expression
$\tau_{s}^{e(IX)}=1/(\tau _{c} \Omega _{SO}^{2})$.
Note also, that both $\tau_c$ and $\tau_0$ depend on the carrier
density and are {\it a priori} different for electrons and IXs in a
CQW.

The longitudinal magnetic field affects differently the spins of
residual electrons and the spins of electrons bound to holes within
IXs.
The precession of the electron spin in the longitudinal magnetic
field averages out the random effective field due to spin-orbit
interaction during the correlation time $\tau _{c}$.
 The spin relaxation time in the presence of the
magnetic field is given by $\tau _{s}^{e(IX)}(B)=(1+(\Omega _{L}\tau
_{c})^{2})\tau _{s}^{e(IX)}$, where $\Omega _{L}$ is
 the Larmor frequency \cite{DyakonovBook}.
For the electrons which are not bound into the excitons and thus are
subject to the cyclotron motion, an additional mechanism further
suppresses the Dyakonov-Perel relaxation \cite{Ivchenko1976}.
In this case, the field induced increase of the spin relaxation time
is given by $\tau _{s}^{e}(B)=(1+(\Omega _{c}\tau _{c})^{2})\tau
_{s}^{e}$, where $\Omega _{c}$ is the cyclotron frequency.
Cyclotron frequency is two orders of magnitude  higher than Larmor
frequency
 so that spin relaxation of the free electrons is expected to be much stronger
 affected by the magnetic field.

Our strategy is  to map both experimentally observed spin components
($\tau_s=250$~ps, $\tau_s(B)/\tau_s=3$  and $\tau_s=10$~ns,
$\tau_s(B)/\tau_s=1$) to this model.
Fig. \ref{FigTau} shows the result of such mapping for fast (a, b)
and slow (c, d) components, assuming either electron (a, c) or IX
(b, d) relaxation.
Pink areas mark the mobility region where $\tau_c >\tau_0$.
Dark lines  show the results of the calculation of the ratio
$\tau_0/ \tau_c$   which fit the experimental results, for each
value of the correlation time.
One can see that in Fig. \ref{FigTau}~(b) there are no dark points
at all.
This means that the fast component is necessarily due to electrons,
mainly because it is strongly affected by the magnetic field.
These electrons are mobile, since $\tau_c > \tau_0$.
By exclusion, the slow component should result from the spin
relaxation of  IX, which are essentially localized $\tau_c <
\tau_0$.
%


%

\end{document}